\documentclass[sn-nature, super]{sn-jnl}

\usepackage{graphicx}%
\usepackage{multirow}%
\usepackage{amsmath,amssymb,amsfonts}%
\usepackage{amsthm}%
\usepackage{mathrsfs}%
\usepackage[title]{appendix}%
\usepackage{xcolor}%
\usepackage{textcomp}%
\usepackage{manyfoot}%
\usepackage{booktabs}%
\usepackage{algorithm}%
\usepackage{algorithmicx}%
\usepackage{algpseudocode}%
\usepackage{listings}%
\usepackage {longtable}

\usepackage{ltablex}
\usepackage{lscape}
\usepackage{setspace}
\usepackage{array}
\usepackage{bibentry}
\usepackage{pgfplots}
\pgfplotsset{compat=1.18}
\usepackage{enumitem}
\setlist[enumerate]{left=0pt, label={[\arabic*]}, labelsep=0.5em}
\usepackage{booktabs,tabularx,ragged2e}
\newcolumntype{Y}{>{\RaggedRight\arraybackslash}X}
\newcolumntype{Z}{>{\RaggedRight\arraybackslash}p{2cm}}
\newcolumntype{Q}{>{\RaggedRight\arraybackslash}p{3.2cm}}

\newcolumntype{P}[1]{>{\centering\arraybackslash}p{#1}}
\newcolumntype{L}[1]{>{\raggedright\arraybackslash}p{#1}}

\newcommand{\tabularxcaption}[2]{%
    \begingroup
    \setbox\tabcapbox\vbox{\tablecaptionfont\raggedright%
    {\bfseries 
    \mbox{#1}\nobreak}{\hskip2mm}\mbox{#2}\vphantom{y}\par\vskip\belowcaptionskip}%
    \box\tabcapbox%
    \endgroup
}
\usepackage{url}

\usepackage{geometry}
\theoremstyle{thmstyleone}%
\theoremstyle{thmstyletwo}%

\theoremstyle{thmstylethree}%

\usepackage{soul}
\usepackage{tikz}
\usetikzlibrary{calc}

\makeatletter
\newif\if@anonymize

\@anonymizefalse  

\if@anonymize
  \newcommand{\highlight@DoHighlight}{
    \fill [outer sep = -15pt, inner sep = 0pt, color=black]
          ($(begin highlight)+(0,8pt)$) rectangle ($(end highlight)+(0,-3pt)$) ;
  }

  \newcommand{\highlight@BeginHighlight}{
    \coordinate (begin highlight) at (0,0) ;
  }

  \newcommand{\highlight@EndHighlight}{
    \coordinate (end highlight) at (0,0) ;
  }

  \newdimen\highlight@previous
  \newdimen\highlight@current
  \newlength{\item@width}

  \DeclareRobustCommand*\anonymize{%
    \SOUL@setup
    \def\SOUL@preamble{%
      \begin{tikzpicture}[overlay, remember picture]
        \highlight@BeginHighlight
        \highlight@EndHighlight
      \end{tikzpicture}%
    }%
    \def\SOUL@postamble{%
      \begin{tikzpicture}[overlay, remember picture]
        \highlight@EndHighlight
        \highlight@DoHighlight
      \end{tikzpicture}%
    }%
    \def\SOUL@everyhyphen{%
      \discretionary{%
        \SOUL@setkern\SOUL@hyphkern
        \SOUL@sethyphenchar
        \tikz[overlay, remember picture] \highlight@EndHighlight ;%
      }{%
      }{%
        \SOUL@setkern\SOUL@charkern
      }%
    }%
    \def\SOUL@everyexhyphen##1{%
      \SOUL@setkern\SOUL@hyphkern
      \settowidth{\item@width}{##1}%
      \makebox[\item@width]{}%
      \discretionary{%
        \tikz[overlay, remember picture] \highlight@EndHighlight ;%
      }{%
      }{%
        \SOUL@setkern\SOUL@charkern
      }%
    }%
    \def\SOUL@everysyllable{%
      \begin{tikzpicture}[overlay, remember picture]
        \path let \p0 = (begin highlight), \p1 = (0,0) in \pgfextra
          \global\highlight@previous=\y0
          \global\highlight@current =\y1
        \endpgfextra (0,0) ;
        \ifdim\highlight@current < \highlight@previous
          \highlight@DoHighlight
          \highlight@BeginHighlight
        \fi
      \end{tikzpicture}%
      \settowidth{\item@width}{\the\SOUL@syllable}%
      \makebox[\item@width]{}%
      \tikz[overlay, remember picture] \highlight@EndHighlight ;%
    }%
    \SOUL@
  }
\else
  \newcommand{\anonymize}[1]{#1}
\fi
\makeatother

\renewcommand{\orcidlogo}{%
  \includegraphics[width=10pt]{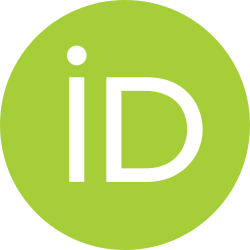}%
}

\begin{document}

\title[From Everyday to Existential]{From Everyday to Existential }
\subtitle{The ethics of shifting the boundaries of health and data with multimodal digital biomarkers }

\author*[1]{\fnm{\anonymize{Joschka}} \sur{\anonymize{Haltaufderheide} \orcid{https://orcid.org/0000-0002-5014-4593}}}\email{\anonymize{joschka.haltaufderheide@uni-potsdam.de}}
\equalcont{\anonymize{Joschka Haltaufderheide} and \anonymize{Florian Funer} contributed equally to this work.}
\author[1]{\fnm{\anonymize{Florian}} \sur{\anonymize{Funer} \orcid{https://orcid.org/0000-0001-9242-0827}}}
\equalcont{\anonymize{Joschka Haltaufderheide} and \anonymize{Florian Funer} contributed equally to this work.}
\author[1]{\fnm{\anonymize{Esther}} \sur{\anonymize{Braun} \orcid{https://orcid.org/0000-0003-4369-7752}}}
\author[1]{\fnm{\anonymize{Hans-Jörg}} \sur{\anonymize{Ehni} \orcid{https://orcid.org/0000-0001-5900-2228}}}
\author[1]{\fnm{\anonymize{Urban}} \sur{\anonymize{Wiesing} \orcid{https://orcid.org/0000-0003-4957-4323}}}
\author[1]{\fnm{\anonymize{Robert}} \sur{\anonymize{Ranisch} \orcid{https://orcid.org/0000-0002-1676-1694}}}

\affil*[1]{\orgdiv{\anonymize{Juniorprofessorship for Medical Ethics with a focus on Digitization}}, \orgname{\anonymize{Faculty for Health Sciences Brandenburg, University of Potsdam}}, \orgaddress{\street{\anonymize{Am M\"uhlenberg 9}}, \city{\anonymize{Potsdam}}, \postcode{\anonymize{14476}}, \state{\anonymize{Brandenburg}}, \country{\anonymize{Germany}}}}

\affil[2]{\orgdiv{\anonymize{Institute for Ethics and History of Medicine}}, \orgname{\anonymize{Medical Faculty,  Eberhard Karls University of Tübingen}}}


 \abstract{Multimodal digital biomarkers (MDBs) integrate diverse physiological, behavioral, and contextual data to provide continuous representations of health. This paper argues that MDBs expand the concept of digital biomarkers along the dimensions of variability, complexity and abstraction, producing an ontological shift that datafies health and an epistemic shift that redefines health relevance. These transformations entail ethical implications for knowledge, responsibility, and governance in data-driven, preventive medicine.}

\keywords{AI, digital biomarkers, prediction, predictive modeling, precision medicine, prevention, conceptual analysis, ethics}



\maketitle
\section*{Background}
Digital biomarkers are often broadly understood as “objective, quantifiable, physiological, and behavioral data that are collected and measured by means of digital devices”\cite{babrakTraditionalDigitalBiomarkers2019, piauCurrentStateDigital2019}.  This definition encompasses a broad range of applications in risk assessment, diagnostics, monitoring, and prediction of health conditions. Digital biomarkers have, for example, been used to detect cognitive decline in healthy individuals at risk of neurodegenerative diseases\cite{ericksonDigitalBiomarkersNeurodegenerative2025} through video games \cite{goldDigitalTechnologiesBiomarkers2018}, computer and smartphone use\cite{dagumDigitalBiomarkersCognitive2018, kayeUnobtrusiveMeasurementDaily2014, liuDigitalPhenotypesMobile2024, parkDiscriminantPowerSmartphoneDerived2024} or assessment of digital navigation behaviours\cite{bayatGPSDrivingDigital2021, deppGPSMobilityDigital2019}, gait measures\cite{manciniDigitalGaitBiomarkers2025} and walking speed\cite{parkDigitalBiomarkersPhysical2021}. In the diagnostic spectrum, digital biomarkers might further aid the diagnosis of conditions such as ADHD\cite{liuAuxiliaryDiagnosisChildren2024, varelacasalClinicalValidationEye2019} or autism spectrum disorders\cite{ponzoAppCharacteristicsAccuracy2023}. In addition, digital biomarkers might prove helpful in monitoring or predicting disease progression\cite{coravosDevelopingAdoptingSafe2019}.
 
While early research on digital biomarkers focused on single data modalities, advances in artificial intelligence and  multimodal data fusion\cite{klineMultimodalMachineLearning2022} have sparked interest in integrating more diverse data sources\cite{powellWalkTalkThink2024, qiAlzheimersDiseaseDigital2025}. This development has given rise to so-called multimodal digital biomarkers (MDBs)—integrative constructs that combine multiple data modalities such as speech patterns, vocal characteristics, and movement data\cite{banksClinicalClassificationMemory2024, chooExploringMultimodalApproach2024, jeongApplicationsDeepLearning2022, qiAlzheimersDiseaseDigital2025}. Recent research on MDBs has demonstrated their potential across numerous health domains, including  neurological disorders such as Alzheimer’s and Parkinson’s disease\cite{jonellMultimodalCapturePatient2021, parkUsingMachineLearning2025, qiAlzheimersDiseaseDigital2025}, mental health assessment\cite{chooExploringMultimodalApproach2024}, as well as the monitoring and management of cardiovascular, respiratory, metabolic, and endocrine conditions\cite{hurwitzHarnessingConsumerWearable2024, paiMultimodalDigitalPhenotyping2024, sfayyihNoninvasiveDiagnosisLung2025}. Data sources may include inertial sensors for gait and movement analysis\cite{pratiharIntegrativeFederatedLearning2025, psaltosMultimodalWearableSensors2019}, audio recordings for acoustic and linguistic evaluation \cite{psaltosMultimodalWearableSensors2019}, keystroke and touchscreen interactions\cite{alfalahiDiagnosticAccuracyKeystroke2022}, physiological measurements, or video data capturing  facial expressions and eye movements\cite{chooExploringMultimodalApproach2024}. By leveraging complementarities among diverse signals, MDBs have shown promising results in terms of  accuracy and reliability in predicting certain health trajectories or outcomes and tailoring interventions to individual needs\cite{banksClinicalClassificationMemory2024, chooExploringMultimodalApproach2024, vairavanMultimodalDigitalBiomarker2023}. 

Despite rapid progress in technical and health research on MDBs, their ethical implications have received little attention so far\cite{andreolettiMappingEthicalLandscape2024}. This may be attributable to the fragmented nature of the field, with related concepts appearing under various terms and in various disciplines such as clinical medicine, health sciences, and technical sciences\cite{maciasalonsoDefinitionsDigitalBiomarkers2024, montagBlurryBoundariesWhen2021}. Therefore, ethical implications of research on MDBs remain underexplored and have yet to be systematically addressed.

Against this background, this paper aims to provide orientation for an ethical discussion on MDBs. We begin by situating MDBs within the conceptual development of digital biomarkers and showing how they expand the concept in terms of variability, complexity, and abstraction. We argue that MDBs present new phenomena with distinctive characteristics. Building on this, we identify two interrelated shifts—the ontological datafication of health and the epistemic recontextualization of data—and trace their ethical implications about the value-laden nature of data and the shift of medicine towards a preventive paradigm. In doing so, we identify potentially new or more pronounced challenges that merit attention as the field advances. We finally discuss implications for research concluding that researchers need to be aware of the distinctive nature of these constructs as well as their ethical implications. In addition to common frameworks of responsible research we suggest that at least three important points need to be considered when engaging with the development of MDBs

\section*{From Single Sources to Multiple Modalities}
Understanding whether MDBs warrant recognition as a distinct phenomenon requires clarifying how they extend the concept of the digital biomarker. Despite increasing usage since around 2014, a unified definition of digital biomarkers remains elusive\cite{maciasalonsoDefinitionsDigitalBiomarkers2024, montagBlurryBoundariesWhen2021, mulinariAligningDigitalBiomarker2024, vasudevanDigitalBiomarkersConvergence2022}. At least seven competing definitions are commonly cited in the literature, reflecting the field’s disciplinary diversity across medicine, informatics, computer science, and engineering\cite{maciasalonsoDefinitionsDigitalBiomarkers2024}.

From a biomedical perspective, digital biomarkers are understood to continue the logic of traditional biomarkers as measurable characteristics indicating normal biological processes, pathological changes, or pharmacological responses\cite{biomarkersdefinitionworkinggroupBiomarkersSurrogateEndpoints2001}. The World Health Organization broadens this notion, describing biomarkers as “any substance or process that can be measured in the body or its products” and that predicts the onset or progression of disease\cite{inter-organizationprogrammeforthesoundmanagementofchemicalsBiomarkersRiskAssessment2001}. Within this framework, biomarkers are understood as medical indicators of healthy or pathological human biology in the broadest sense—observable but not necessarily subjectively perceptible symptoms\cite{strimbuWhatAreBiomarkers2010}—and are distinct from clinical endpoints, which refer to direct measures of patient outcomes or well-being.

Digital biomarkers, in this perspective, retain this foundational logic—a biological or physiological substrate, a measurable proxy, and a method of quantification—but differ in their mode of data acquisition. The term “digital” typically refers to the use of smartphones, wearables, ambient sensors, or other networked consumer devices that facilitate non-invasive, low-cost, and continuous data collection in real-world environments\cite{babrakTraditionalDigitalBiomarkers2019, coravosDevelopingAdoptingSafe2019}. This transition is thought to overcome many limitations traditionally associated with biomarker acquisition, such as episodic measurement or procedural invasiveness\cite{babrakTraditionalDigitalBiomarkers2019, hartlTranslationalPrecisionMedicine2021, maciasalonsoDefinitionsDigitalBiomarkers2024}.

Complementing the biomedical lens, a perspective rooted in data science and engineering foregrounds that what makes biomarkers “digital” lies less in the devices used than in the computational transformation of data\cite{coravosDevelopingAdoptingSafe2019, maciasalonsoDefinitionsDigitalBiomarkers2024}. Digital biomarkers are typically derived from high-volume, unstructured, and passively collected biological, physiological, and sometimes behavioral data, which are processed through algorithmic pipelines\cite{dorseyFirstFrontierDigital2017} to yield diagnostic and predictive insights. 

Against the background of these complementing perspectives, each revealing crucial aspects of the concept of a digital biomarker, the emergence of MDBs can be understood as both an evolution and an expansion of the digital biomarker concept along three interrelated dimensions:  (i) variability, (ii) complexity and (iii) level of abstraction. An overview is given in Fig. 1.

\begin{enumerate}[label=\roman*., leftmargin=*]
    \item \textbf{Variability:} MDBs integrate  multiple data types within a unified analytic framework. This multimodality, it is assumed, enables a more nuanced and context-sensitive representation of health and disease, potentially capturing phenomena that unimodal approaches may overlook. This entails a growing consideration of data modalities that were previously not regarded as sources for biomarkers like, for instance, behavioral and self-reported data that has been contested as a credible source for biomarkers\cite{montagBlurryBoundariesWhen2021, mulinariAligningDigitalBiomarker2024}. 

    \item \textbf{Complexity:} The fusion of heterogeneous data types introduces substantial analytical and computational complexity. Advanced machine learning, particularly multimodal fusion, is required to manage data dimensionality and to model intricate interdependencies across modalities to extract meaningful features. Complexity thus refers both to the inherent challenges of analyzing vast and diverse datasets as well as to the rationales embedded in the algorithms that process such data43. The resulting outputs can be classified as “inferential data”\cite{wachterRightReasonableInferences2019}—that is, information statistically derived  from observable data points that individually bear no evident or direct relation to the represented health attribute.

    \item \textbf{Abstraction:} MDBs heighten abstraction in two respects: First, they increasingly detach from the human body as the primary ontological reference for biomedical data\cite{vayenaStrictlyBiomedicalSketching2016}, since MDBs integrate data streams whose connection is often indirect, mediated, or computationally inferred. Second, they move beyond perceivable symptoms by detecting subtle, often imperceptible alterations. Certain data types, insignificant in isolation, gain significance only through their inferential integration into a multimodal analytic framework.
\end{enumerate}

\begin{figure}[t]  
\centering
\caption{\label{fig1:expansion}The gradual expansion of the biomarker concept}
\includegraphics[width=\textwidth]{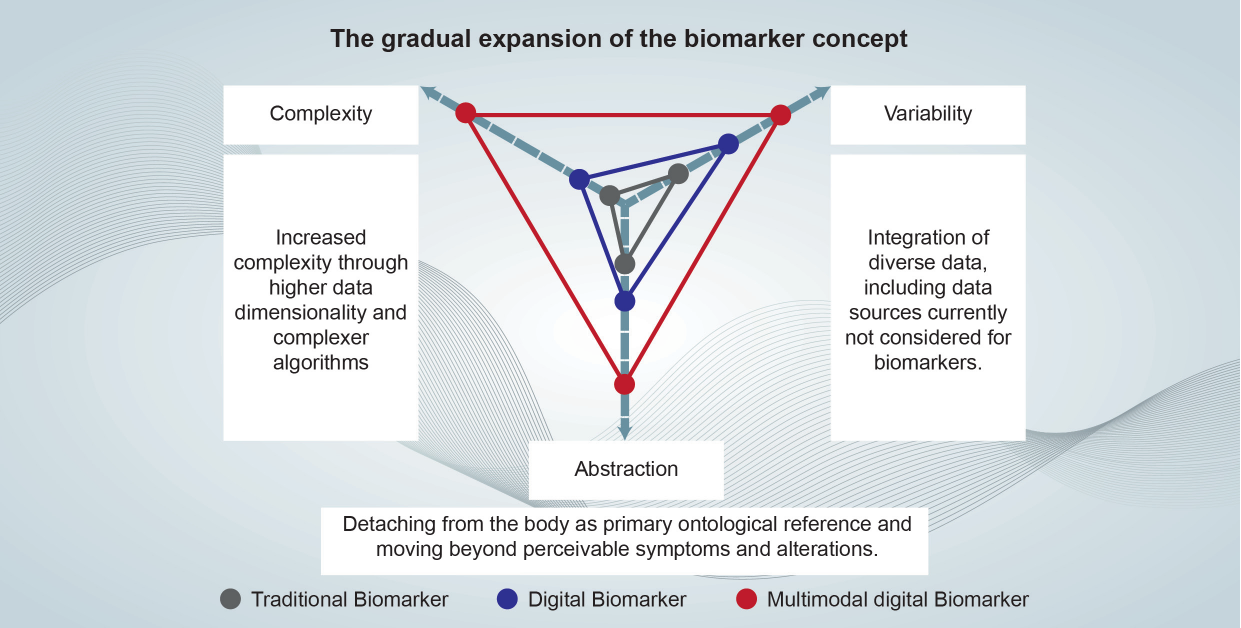}
\end{figure}

\section*{Expanding the Paradigm — Expanding the Ethical Implications}
Consequently, MDBs can be regarded as a conceptual and epistemic expansion of digital biomarkers. They redefine how health is represented, inferred, and rendered actionable—transformations that, in turn, may carry profound ethical implications.

To assess the ethical significance of these transformations, we propose to understand the expansion along two complementary, yet mutually reinforcing directions. Borrowing from philosophy of science, we describe these as an ontological shift, which redefines what health is, and an epistemic shift, which reshapes what counts as health-relevant.
 
\subsection*{The Ontological Shift: Datafication of Health and Disease} 
On the one hand, MDBs exemplify the ongoing process of datafication of health and disease—the translation of bodily, behavioral, and social phenomena into digital data that can be captured, computed, and optimized\cite{ruckensteinDataficationHealth2017, vandijckDataficationDataismDataveillance2014}. Datafication can be understood as the process of transforming of aspects of human life into digital data\cite{mejiasDatafication2019}. However, rather than a mere quantification, it encompasses a process that is structured to generate informational (and economic) value. Datafication creates knowledge that can be operationally integrated into decision-making and thus becomes actionable within computational infrastructures. In MDBs, health or disease are no longer primarily observable or biological indicators of the body (as originally indicated by the term biomarker) but statistical constellations of data points. What emerges from this transformation is, hence, more than an informational artifact but an information that matters for health decisions. MDBs, in this sense, contribute to concepts of health and disease becoming actionable as a signature within larger data infrastructures, involving a reframing of what health or disease is. Normatively, this demands reflection on how far such computational ontologies should guide diagnosis, prediction, and intervention.

\subsection*{The Epistemic Shift: Recontextualization or Healthization of Data}
On the other hand, the extension of the digital biomarker concept through MDBs drives a process of recontextualization that may also be described as healthization of data. Through multimodal integration, data sources previously not regarded as medically relevant or health-related—such as interactional habits or mobility patterns—are reframed as relevant to health indicators once algorithmically connected to physiological or behavioral states. Recontextualisation of data can be understood as the process by which everyday data types or data emitting practices gain medical and moral significance because of their potential to contribute to health-related decisions in risk assessment, diagnosis, therapy, monitoring or prevention. Health relevance, therefore, is no longer discovered, it is constructed through linkage\cite{mejiasDatafication2019, ruckensteinDataficationHealth2017}. Hence, the epistemic redefinition of what is knowable about health reshapes the moral landscape of how individuals and institutions are expected to act upon it.

\begin{table}[htbp]
\caption{The two directions of the expansion of biomarkers. }
\begin{tabularx}{\linewidth}{@{}Q Y Z Y@{}}
\toprule
\textbf{Expanding direction} & \textbf{Definition} & \textbf{Type of shift} & \textbf{Key normative implication} \\
\midrule
Datafication of health &
Translation of physiological, behavioral, and social phenomena into digital data constituting health as a data object &
Ontological (what health is) &
Calls for ethical reflection on the legitimacy and limits of acting on computational representations of health \\
\addlinespace
Recontextualization / Healthization of data &
Algorithmic transformation of non-medical data into health-relevant information &
Epistemic (what counts as health-relevant) &
Expands the moral and social expectations attached to health-related behavior and data use \\
\bottomrule
\end{tabularx}
\end{table}

This dual dynamic blurs distinctions between clinical and everyday aspects. While not unique to MDBs, their intensified variability, complexity, and abstraction amplify this effect, expanding ethical inquiry far beyond traditional concerns such as privacy or data accuracy. Its focus must therefore include how MDBs reshape knowledge production and the moral expectations attached to datafied health. In what follows, we unpack some of the normative implications along two ethical anchoring points: 1.) the production of data in the context of health; and 2.) the effects of data on the paradigm of prevention and the moral imperatives of datafied medicine. 

\subsection*{MDBs and the Implications of Producing Health Data }
As part of the datafication of health, central ethical issues in relation to MDBs align with broader concerns in data and AI ethics. Key issues arise from how data are produced and their outcome is validated as well as from the ability to infer health-relevant information from sources lacking an obvious link to the measured phenomena.  

Critical scholarship has emphasized that data generation—by users who emit data—as well as data production—via the compilation of data from multiple sources—do not resemble the neutral representation of facts\cite{hoeyerDataficationAccountabilityPublic2019, luptonCriticalDigitalHealth2016, ruckensteinDataficationHealth2017}. Rather, it constitutes a complex, socially situated phenomenon that is inherently value-laden. Decisions about who and how something is represented, and how analytical relationships are constructed shape the resulting MDB. With each additional step, the data undergoes transformations by passing through an interpretive framework\cite{mittelstadtEthicsBigData2016}. Unlike the term biomarker originally  suggests, MDBs fuse data sources in ways that reflect not only the physical world but also its social fabric, that is, the normative structures and processes through which they are created. Consequently, the validity and reliability of MDBs cannot be understood as purely technical or methodological properties, but are contingent upon the social contexts in which data are generated, selected, refined and interpreted. What counts as data, whose data are included, how variation is classified, and how data are processed define the epistemic limits of what can meaningfully represent health. Ethical considerations therefore extend beyond privacy or data control to include questions of epistemic justice and fairness: For example, social and demographic biases inscribed in data infrastructures may systematically privilege certain body types, behaviors, or life-worlds while marginalizing others. MDBs thus not only reflect existing inequities but may amplify them, reinforcing structural disparities in whose health can be rendered visible and actionable through data-driven means.

Particular ethical attention is needed because MDB outputs are inferential. As has been argued in discussions about big data, data-driven inferences may introduce a complex and profound informational asymmetry of knowledge and control. With the recontextualisation of health data and the inferential nature of MDBs, individuals typically are unaware of what could be extrapolated from the data they emit, and they typically remain unaware when, to whom or by whom this information can be revealed. While the revealed information might be worthy of protection from an ethical perspective, the source it was inferred from might not be. Recontextualization of data, hence, shifts the challenge of being able to control the distribution of personal information towards the challenge of being able to control the use of data\cite{vayenaStrictlyBiomedicalSketching2016} while its significance may change over time. This would require new practices of government and control spanning the whole life-cycle of data—however Wachter and Mittelstadt have shown that the status of inferred or derived data currently remains ambiguous in most legal data protection regimes. Because such data are not directly provided by the data subject but rather produced through algorithmic processing, they often fall outside existing rights of access, correction, or contestation. This unresolved legal position amplifies the ethical stakes, as individuals are not only deprived of practical control but also lack formal means to challenge or even know about the inferences made about them\cite{wachterRightReasonableInferences2019}.

\subsection*{MDBs, Anticipatory Medicine, and the Moral Responsibilization of Health }
The ontological and epistemic transformations introduced by MDBs also entail a distinct moral and practical reconfiguration of medicine. If datafication renders health as a continuous informational state and recontextualization extends what counts as health-relevant data, MDBs not only transform how health is known but also how health decisions are made. In this regard, the discourse on MDBs is connected to longstanding debates on the expansion of medicine. As Hoffmann has argued, medicine tends to extend its boundaries along multiple dimensions—conceptual, social, and technological—by redefining what counts as disease, who qualifies as a patient, and what becomes subject to medical scrutiny\cite{hofmannManagingMoralExpansion2022}. Such expansions occurs not merely through clinical discovery but through the adoption of new epistemic tools and ontological shifts that broaden the scope of what medicine can see and act upon. Medicine thereby increasingly moves from the “present to the future”, acting pre-emptively through prediction and risk management; from the “normal to the abnormal”, redefining subtle variations as early pathology and  from the “body to life”, extending medical concern to behavioral and social domains. 

MDBs exemplify this multidirectional expansion: Deviations in digital signatures can become interpretable as early or pre-symptomatic signs of pathology. With this, MDBs merge diagnosis with prediction, turning potential futures into actionable presents. The question of the right time for a diagnosis, the kairos in diagnostics\cite{hofmannKairosDiagnostics2024}, is not even asked in this context; rather, it simply happens with the use of MDBs. Furthermore, they extend attention beyond the body into everyday life, for example, into the ways people type, walk, speak, or interact, transforming ordinary activities into continuous sources of health-relevant information, thereby expanding the spectrum of what medicine can problematize and address. This reconfiguration risks creating individuals who become patients-in-waiting, suspended between well-being and disease, monitored for potential deviations that may never manifest clinically. 

These processes resonate with different notions of pathologization and healthization—the extension of disease and health categories into ever broader aspects of human life—as well as stronger critics of (bio-)medicalization that view the growing technologization of health as a process of moral and social normalization with which individuals are expected not only to avoid illness but to actively pursue and optimize health as a continuous project. 

The normative implications of these transformations are significant. MDBs advance a “preventive paradigm” that reframes medicine as an anticipatory, ongoing and self-regulatory enterprise. On the one hand, self-regulation, is intimately linked to the notion of autonomy, suggesting an expansion of individuals’ capacity to know and act through new forms of digitally mediated self-reflexivity. However, as predictive inferences are translated into health-relevant knowledge, preventive health maintenance is also recast as a standing personal responsibility. Individuals are invited—or subtly nudged—to monitor, interpret, and optimize their own data traces to maintain a favorable digital health profile as an exercise of this duty. Rose describes this development as part of a new ethico-politics of vitality, in which individuals are expected to manage their biological existence responsibly as “biological citizens”\cite{roseBiologicalCitizenship2005, rosePoliticsLifeItself2009}. Health becomes a matter of moral obligation\cite{brownMoralResponsibilisationHealth2019}: continuous self-surveillance and data-informed decision-making signify good and responsible citizenship. As critical work on digital health shows, the expansion of knowing and acting often coincides with this redistribution of responsibility, blurring the line between empowerment on the one hand and responsibilization—the transfer of moral and practical responsibility for health maintenance from institutions to individuals beyond what is reasonable and fair—on the other. Because MDBs operationalize anticipatory knowledge through multimodal signatures they can intensify both sides of this relation.

In addition, as MDBs are processed within algorithmic pipelines\cite{dorseyFirstFrontierDigital2017}, they normalize an ethos of data-driven self-care, where deviation from modeled MDB norms may appear as risk, and risk becomes a site of intervention. The resulting biopolitical logic distributes health obligations unevenly: those with limited access to digital technologies or data literacy may find themselves excluded or problematized as less compliant, less responsible, or even less deserving. MDBs, in this sense, not only extend the predictive and diagnostic reach of medicine but also produce new hierarchies of normality, meaning that what is considered normal, healthy or desirable is increasingly defined by how individuals register in data. This goes along with new expectations of what it means to be a responsible patient and citizen in data-intensive healthcare systems, mainly in situations where symptoms are not (yet) perceivable.

\section*{Implications for research}
With this analysis, we reveal that MDBs are associated with a range of ethical challenges. They gradually extend the biomarker concept beyond traditional data sources, increase analytical complexity through inferential rather than observable information, and shift attention from experienced to non-experienced phenomena and paraclinical indicators\cite{hofmannManagingMoralExpansion2022}. MDBs are therefore not only predictive or diagnostic tools but normative technologies that reshape how health, disease and related moral judgments are defined and distributed. Fig. 2 schematizes the underlying mechanics of datafication and recontextualization.

\begin{figure}[h!]  
\centering
\caption{\label{fig1:prisma}Datafication and recontextualisation}
\includegraphics[width=\textwidth]{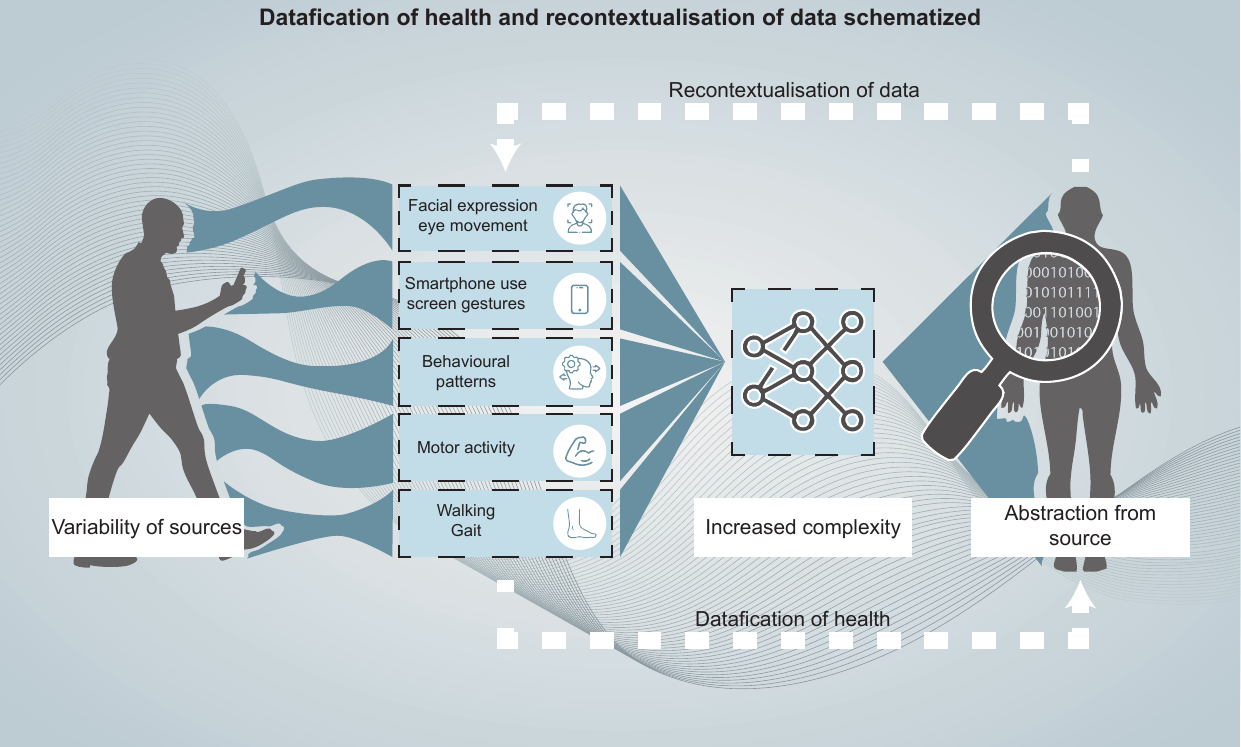}
\end{figure}

As research on MDBs accelerates, supported by emerging evidence on promising benefits, their distinctive characteristics demand careful consideration. Given the challenges described in this paper, their further development requires responsible research practices that include ethical considerations from the beginning. Research on data-intensive health practices such as MDBs does in fact require more human intelligence and meaning-making, the more computational power and data is put to use\cite{mittelstadtEthicsBigData2016}. Accordingly, we suggest that responsible research and development of MDBs should include an ex-ante process of ethical consideration, centering around three guiding questions: 

\begin{enumerate}[label=\roman*., leftmargin=*]
    \item Why might certain data provide a normatively acceptable basis for MDB development? 

    \item Why should the inferences that could be drawn from this data be regarded as relevant from a perspective of good medicine?

    \item Can the intended outcome be achieved, based on an assessment of their validity and reliability?
\end{enumerate}

The first question seeks to avoid misconceiving data as an objective analogue to reality\cite{mittelstadtEthicsBigData2016}, shedding light on what is relevant and what the actual purpose of an investigation towards a potential MDB might be. It highlights the importance of understanding data in its social context, both where it originates and where it will be used. This requires assessing the specific risk profile of the data used in comparison to potential alternatives and clarifying the epistemic and practical purpose of developing an MDB. 

The second question considers potential effects of inferential and complex data in light of the aims of good medicine. Good medicine prima facie prioritizes pain and suffering over non-experienced indicators, present over future states, and relieving harm over pursuing positive well-being\cite{hofmannManagingMoralExpansion2022}. It also entails fairness, understood as a just distribution of goods and opportunities for participation. Although justified exceptions are possible, they bear the burden of proof—for example, by showing that an MDB could genuinely improve medical practice\cite{hofmannManagingMoralExpansion2022}. Thus, answering the second question requires moving beyond assumed advantages of MDBs (e.g., overcoming current limitations or simplifying data access) to articulate concrete, defensible aims consistent with the goals of good medicine. 
The third question evaluates whether the anticipated benefits of MDBs are realistically achievable, given their empirical validity, reliability, and contextual feasibility. This requires a critical comparison of expected gains with foreseeable risks.

However, given the unpredictable effects that result from data recontextualization as well as potentially irreversible societal effects, this process needs to be bolstered by robust mechanisms to ensure ethical research. Because the full impact of MDB research and deployment will only unfold over time and once its output takes shape, it may be fruitful to consider this type of research as a social experiment as has been suggested recently\cite{ranischFoundationModelsMedicine2025}. When advanced information technologies are introduced into healthcare, society or parts of it effectively become a “real-world laboratory” in which experimental interventions, uncertain outcomes, and variable user interactions take place. Applied to research on MDBs, this framing implies that researchers should not assume that aims are self-justifying or effects predictable. Instead, MDB research should embed oversight, gradual roll-out, continuous monitoring, iterative learning, and transparency to prevent large-scale harms, hidden biases, or systemic inequities from emerging only after deployment. Such a framing demands that MDB research includes accompanying ethical evaluation, adaptive governance, and explicitly defined mechanisms for identifying, correcting, or halting harmful developments throughout the research cycle.

\subsection*{Conclusions}
MDBs signify a turning point in the evolution of data-intensive medicine. They expand the biomarker concept from single-modality observation to multimodal inference, transforming health into a dynamic constellation of data relations. The ethical implications of MDBs therefore reach beyond familiar debates and concern the very conditions under which health becomes knowable, actionable and morally charged. With this, MDBs call for a new form of ethical vigilance attuned to how data infrastructures reshape both medical knowledge and the moral imperatives of care.

\section*{Statements and Declarations}
\subsection*{Acknowledgements}
\anonymize{Open Access funding enabled and organized by Projekt DEAL. This work was funded by the VolkswagenStiftung as part of the Digital Medical Ethics Network (Funding No. 9B 233).} Fig 1. was created with material from juicy fish on Freepik. Fig 2. was created with material from juicy fish on Freepik and Goodware, freepik, Anatoly, pojok d and Those Icons from flaticon.

\subsection*{Author contributions}

\anonymize{J.H. and F.F. conceived the manuscript, drafted different sections of it, led the drafting process, and coordinated revisions. E.B., H.-J.E., U.W., and R.R. contributed to the conceptual development and revised several drafts.} All authors read and approved the final manuscript.

\subsection*{Competing interests}

The authors declare no competing interests.


\begin{thebibliography}{10}
\expandafter\ifx\csname url\endcsname\relax
  \def\url#1{\burl{#1}}\fi
\expandafter\ifx\csname urlprefix\endcsname\relax\def\urlprefix{URL }\fi
\providecommand{\bibinfo}[2]{#2}
\providecommand{\eprint}[2][]{\url{#2}}
\providecommand{\doi}[1]{\url{https://doi.org/#1}}
\bibcommenthead

\bibitem{babrakTraditionalDigitalBiomarkers2019}
\bibinfo{author}{Babrak, L.~M.} \emph{et~al.}
\newblock \bibinfo{title}{Traditional and {{Digital Biomarkers}}: {{Two Worlds
  Apart}}?}
\newblock \emph{\bibinfo{journal}{Digital Biomarkers}}
  \textbf{\bibinfo{volume}{3}}, \bibinfo{pages}{92--102}
  (\bibinfo{year}{2019}).

\bibitem{piauCurrentStateDigital2019}
\bibinfo{author}{Piau, A.}, \bibinfo{author}{Wild, K.},
  \bibinfo{author}{Mattek, N.} \& \bibinfo{author}{Kaye, J.}
\newblock \bibinfo{title}{Current {{State}} of {{Digital Biomarker
  Technologies}} for {{Real-Life}}, {{Home-Based Monitoring}} of {{Cognitive
  Function}} for {{Mild Cognitive Impairment}} to {{Mild Alzheimer Disease}}
  and {{Implications}} for {{Clinical Care}}: {{Systematic Review}}}.
\newblock \emph{\bibinfo{journal}{Journal of Medical Internet Research}}
  \textbf{\bibinfo{volume}{21}}, \bibinfo{pages}{e12785}
  (\bibinfo{year}{2019}).

\bibitem{ericksonDigitalBiomarkersNeurodegenerative2025}
\bibinfo{author}{Erickson, C.~M.}, \bibinfo{author}{Wexler, A.} \&
  \bibinfo{author}{Largent, E.~A.}
\newblock \bibinfo{title}{Digital {{Biomarkers}} for {{Neurodegenerative
  Disease}}}.
\newblock \emph{\bibinfo{journal}{JAMA Neurology}}
  \textbf{\bibinfo{volume}{82}}, \bibinfo{pages}{5} (\bibinfo{year}{2025}).

\bibitem{goldDigitalTechnologiesBiomarkers2018}
\bibinfo{author}{Gold, M.} \emph{et~al.}
\newblock \bibinfo{title}{Digital technologies as biomarkers, clinical outcomes
  assessment, and recruitment tools in {{Alzheimer}}'s disease clinical
  trials}.
\newblock \emph{\bibinfo{journal}{Alzheimer's \& Dementia: Translational
  Research \& Clinical Interventions}} \textbf{\bibinfo{volume}{4}},
  \bibinfo{pages}{234--242} (\bibinfo{year}{2018}).

\bibitem{dagumDigitalBiomarkersCognitive2018}
\bibinfo{author}{Dagum, P.}
\newblock \bibinfo{title}{Digital biomarkers of cognitive function}.
\newblock \emph{\bibinfo{journal}{npj Digital Medicine}}
  \textbf{\bibinfo{volume}{1}}, \bibinfo{pages}{10} (\bibinfo{year}{2018}).

\bibitem{kayeUnobtrusiveMeasurementDaily2014}
\bibinfo{author}{Kaye, J.} \emph{et~al.}
\newblock \bibinfo{title}{Unobtrusive measurement of daily computer use to
  detect mild cognitive impairment}.
\newblock \emph{\bibinfo{journal}{Alzheimer's \& Dementia}}
  \textbf{\bibinfo{volume}{10}}, \bibinfo{pages}{10--17}
  (\bibinfo{year}{2014}).

\bibitem{liuDigitalPhenotypesMobile2024}
\bibinfo{author}{Liu, Q.} \emph{et~al.}
\newblock \bibinfo{title}{Digital {{Phenotypes}} of {{Mobile Keyboard Backspace
  Rates}} and {{Their Associations With Symptoms}} of {{Mood Disorder}}:
  {{Algorithm Development}} and {{Validation}}}.
\newblock \emph{\bibinfo{journal}{Journal of Medical Internet Research}}
  \textbf{\bibinfo{volume}{26}}, \bibinfo{pages}{e51269}
  (\bibinfo{year}{2024}).

\bibitem{parkDiscriminantPowerSmartphoneDerived2024}
\bibinfo{author}{Park, J.-H.}
\newblock \bibinfo{title}{Discriminant {{Power}} of {{Smartphone-Derived
  Keystroke Dynamics}} for {{Mild Cognitive Impairment Compared}} to a
  {{Neuropsychological Screening Test}}: {{Cross-Sectional Study}}}.
\newblock \emph{\bibinfo{journal}{Journal of Medical Internet Research}}
  \textbf{\bibinfo{volume}{26}}, \bibinfo{pages}{e59247}
  (\bibinfo{year}{2024}).

\bibitem{bayatGPSDrivingDigital2021}
\bibinfo{author}{Bayat, S.} \emph{et~al.}
\newblock \bibinfo{title}{{{GPS}} driving: A digital biomarker for preclinical
  {{Alzheimer}} disease}.
\newblock \emph{\bibinfo{journal}{Alzheimer's Research \& Therapy}}
  \textbf{\bibinfo{volume}{13}}, \bibinfo{pages}{115} (\bibinfo{year}{2021}).

\bibitem{deppGPSMobilityDigital2019}
\bibinfo{author}{Depp, C.~A.} \emph{et~al.}
\newblock \bibinfo{title}{{{GPS}} mobility as a digital biomarker of negative
  symptoms in schizophrenia: A case control study}.
\newblock \emph{\bibinfo{journal}{npj Digital Medicine}}
  \textbf{\bibinfo{volume}{2}}, \bibinfo{pages}{108} (\bibinfo{year}{2019}).

\bibitem{manciniDigitalGaitBiomarkers2025}
\bibinfo{author}{Mancini, M.} \emph{et~al.}
\newblock \bibinfo{title}{Digital gait biomarkers in {{Parkinson}}'s disease:
  Susceptibility/risk, progression, response to exercise, and prognosis}.
\newblock \emph{\bibinfo{journal}{npj Parkinson's Disease}}
  \textbf{\bibinfo{volume}{11}}, \bibinfo{pages}{51} (\bibinfo{year}{2025}).

\bibitem{parkDigitalBiomarkersPhysical2021}
\bibinfo{author}{Park, C.}, \bibinfo{author}{Mishra, R.},
  \bibinfo{author}{Golledge, J.} \& \bibinfo{author}{Najafi, B.}
\newblock \bibinfo{title}{Digital {{Biomarkers}} of {{Physical Frailty}} and
  {{Frailty Phenotypes Using Sensor-Based Physical Activity}} and {{Machine
  Learning}}}.
\newblock \emph{\bibinfo{journal}{Sensors}} \textbf{\bibinfo{volume}{21}},
  \bibinfo{pages}{5289} (\bibinfo{year}{2021}).

\bibitem{liuAuxiliaryDiagnosisChildren2024}
\bibinfo{author}{Liu, Z.} \emph{et~al.}
\newblock \bibinfo{title}{Auxiliary {{Diagnosis}} of {{Children With
  Attention-Deficit}}/{{Hyperactivity Disorder Using Eye-Tracking}} and
  {{Digital Biomarkers}}: {{Case-Control Study}}}.
\newblock \emph{\bibinfo{journal}{JMIR mHealth and uHealth}}
  \textbf{\bibinfo{volume}{12}}, \bibinfo{pages}{e58927}
  (\bibinfo{year}{2024}).

\bibitem{varelacasalClinicalValidationEye2019}
\bibinfo{author}{Varela~Casal, P.} \emph{et~al.}
\newblock \bibinfo{title}{Clinical {{Validation}} of {{Eye Vergence}} as an
  {{Objective Marker}} for {{Diagnosis}} of {{ADHD}} in {{Children}}}.
\newblock \emph{\bibinfo{journal}{Journal of Attention Disorders}}
  \textbf{\bibinfo{volume}{23}}, \bibinfo{pages}{599--614}
  (\bibinfo{year}{2019}).

\bibitem{ponzoAppCharacteristicsAccuracy2023}
\bibinfo{author}{Ponzo, S.} \emph{et~al.}
\newblock \bibinfo{title}{App {{Characteristics}} and {{Accuracy Metrics}} of
  {{Available Digital Biomarkers}} for {{Autism}}: {{Scoping Review}}}.
\newblock \emph{\bibinfo{journal}{JMIR mHealth and uHealth}}
  \textbf{\bibinfo{volume}{11}}, \bibinfo{pages}{e52377}
  (\bibinfo{year}{2023}).

\bibitem{coravosDevelopingAdoptingSafe2019}
\bibinfo{author}{Coravos, A.}, \bibinfo{author}{Khozin, S.} \&
  \bibinfo{author}{Mandl, K.~D.}
\newblock \bibinfo{title}{Developing and adopting safe and effective digital
  biomarkers to improve patient outcomes}.
\newblock \emph{\bibinfo{journal}{npj Digital Medicine}}
  \textbf{\bibinfo{volume}{2}}, \bibinfo{pages}{14} (\bibinfo{year}{2019}).

\bibitem{klineMultimodalMachineLearning2022}
\bibinfo{author}{Kline, A.} \emph{et~al.}
\newblock \bibinfo{title}{Multimodal machine learning in precision health:
  {{A}} scoping review}.
\newblock \emph{\bibinfo{journal}{npj Digital Medicine}}
  \textbf{\bibinfo{volume}{5}}, \bibinfo{pages}{171} (\bibinfo{year}{2022}).

\bibitem{powellWalkTalkThink2024}
\bibinfo{author}{Powell, D.}
\newblock \bibinfo{title}{Walk, talk, think, see and feel: Harnessing the power
  of digital biomarkers in healthcare}.
\newblock \emph{\bibinfo{journal}{npj Digital Medicine}}
  \textbf{\bibinfo{volume}{7}}, \bibinfo{pages}{45} (\bibinfo{year}{2024}).

\bibitem{qiAlzheimersDiseaseDigital2025}
\bibinfo{author}{Qi, W.} \emph{et~al.}
\newblock \bibinfo{title}{Alzheimer's disease digital biomarkers
  multidimensional landscape and {{AI}} model scoping review}.
\newblock \emph{\bibinfo{journal}{npj Digital Medicine}}
  \textbf{\bibinfo{volume}{8}}, \bibinfo{pages}{366} (\bibinfo{year}{2025}).

\bibitem{banksClinicalClassificationMemory2024}
\bibinfo{author}{Banks, R.} \emph{et~al.}
\newblock \bibinfo{title}{Clinical classification of memory and cognitive
  impairment with multimodal digital biomarkers}.
\newblock \emph{\bibinfo{journal}{Alzheimer's \& Dementia: Diagnosis,
  Assessment \& Disease Monitoring}} \textbf{\bibinfo{volume}{16}},
  \bibinfo{pages}{e12557} (\bibinfo{year}{2024}).

\bibitem{chooExploringMultimodalApproach2024}
\bibinfo{author}{Choo, M.} \emph{et~al.}
\newblock \bibinfo{title}{Exploring a multimodal approach for utilizing digital
  biomarkers for childhood mental health screening}.
\newblock \emph{\bibinfo{journal}{Frontiers in Psychiatry}}
  \textbf{\bibinfo{volume}{15}}, \bibinfo{pages}{1348319}
  (\bibinfo{year}{2024}).

\bibitem{jeongApplicationsDeepLearning2022}
\bibinfo{author}{Jeong, H.} \emph{et~al.}
\newblock \bibinfo{title}{Applications of deep learning methods in digital
  biomarker research using noninvasive sensing data}.
\newblock \emph{\bibinfo{journal}{DIGITAL HEALTH}}
  \textbf{\bibinfo{volume}{8}}, \bibinfo{pages}{205520762211366}
  (\bibinfo{year}{2022}).

\bibitem{jonellMultimodalCapturePatient2021}
\bibinfo{author}{Jonell, P.} \emph{et~al.}
\newblock \bibinfo{title}{Multimodal {{Capture}} of {{Patient Behaviour}} for
  {{Improved Detection}} of {{Early Dementia}}: {{Clinical Feasibility}} and
  {{Preliminary Results}}}.
\newblock \emph{\bibinfo{journal}{Frontiers in Computer Science}}
  \textbf{\bibinfo{volume}{3}}, \bibinfo{pages}{642633} (\bibinfo{year}{2021}).

\bibitem{parkUsingMachineLearning2025}
\bibinfo{author}{Park, H.} \emph{et~al.}
\newblock \bibinfo{title}{Using machine learning to identify {{Parkinson}}'s
  disease severity subtypes with multimodal data}.
\newblock \emph{\bibinfo{journal}{Journal of NeuroEngineering and
  Rehabilitation}} \textbf{\bibinfo{volume}{22}}, \bibinfo{pages}{126}
  (\bibinfo{year}{2025}).

\bibitem{hurwitzHarnessingConsumerWearable2024}
\bibinfo{author}{Hurwitz, E.} \emph{et~al.}
\newblock \bibinfo{title}{Harnessing {{Consumer Wearable Digital Biomarkers}}
  for {{Individualized Recognition}} of {{Postpartum Depression Using}} the
  {{All}} of {{Us Research Program Data Set}}: {{Cross-Sectional Study}}}.
\newblock \emph{\bibinfo{journal}{JMIR mHealth and uHealth}}
  \textbf{\bibinfo{volume}{12}}, \bibinfo{pages}{e54622}
  (\bibinfo{year}{2024}).

\bibitem{paiMultimodalDigitalPhenotyping2024}
\bibinfo{author}{Pai, A.} \emph{et~al.}
\newblock \bibinfo{title}{Multimodal digital phenotyping of diet, physical
  activity, and glycemia in {{Hispanic}}/{{Latino}} adults with or at risk of
  type 2 diabetes}.
\newblock \emph{\bibinfo{journal}{npj Digital Medicine}}
  \textbf{\bibinfo{volume}{7}}, \bibinfo{pages}{7} (\bibinfo{year}{2024}).

\bibitem{sfayyihNoninvasiveDiagnosisLung2025}
\bibinfo{author}{Sfayyih, A.~H.}, \bibinfo{author}{Sulaiman, N.} \&
  \bibinfo{author}{Sabry, A.~H.}
\newblock \bibinfo{title}{Non-invasive diagnosis of lung diseases via
  multimodal feature extraction from breathing audio and chest dynamics}.
\newblock \emph{\bibinfo{journal}{Computers in Biology and Medicine}}
  \textbf{\bibinfo{volume}{191}}, \bibinfo{pages}{110182}
  (\bibinfo{year}{2025}).

\bibitem{pratiharIntegrativeFederatedLearning2025}
\bibinfo{author}{Pratihar, R.} \& \bibinfo{author}{Sankar, R.}
\newblock \bibinfo{title}{Integrative {{Federated Learning Framework}} for
  {{Multimodal Parkinson}}'s {{Disease Biomarker Fusion}}}.
\newblock \emph{\bibinfo{journal}{Computers}} \textbf{\bibinfo{volume}{14}},
  \bibinfo{pages}{388} (\bibinfo{year}{2025}).

\bibitem{psaltosMultimodalWearableSensors2019}
\bibinfo{author}{Psaltos, D.} \emph{et~al.}
\newblock \bibinfo{title}{Multimodal {{Wearable Sensors}} to {{Measure Gait}}
  and {{Voice}}}.
\newblock \emph{\bibinfo{journal}{Digital Biomarkers}}
  \textbf{\bibinfo{volume}{3}}, \bibinfo{pages}{133--144}
  (\bibinfo{year}{2019}).

\bibitem{alfalahiDiagnosticAccuracyKeystroke2022}
\bibinfo{author}{Alfalahi, H.} \emph{et~al.}
\newblock \bibinfo{title}{Diagnostic accuracy of keystroke dynamics as digital
  biomarkers for fine motor decline in neuropsychiatric disorders: A systematic
  review and meta-analysis}.
\newblock \emph{\bibinfo{journal}{Scientific Reports}}
  \textbf{\bibinfo{volume}{12}}, \bibinfo{pages}{7690} (\bibinfo{year}{2022}).

\bibitem{vairavanMultimodalDigitalBiomarker2023}
\bibinfo{author}{Vairavan, S.} \emph{et~al.}
\newblock \bibinfo{title}{A multimodal digital biomarker of functional deficits
  in early-stage {{Alzheimer}}'s disease: Results of the {{RADAR}}-{{AD}}
  study}.
\newblock \emph{\bibinfo{journal}{Alzheimer's \& Dementia}}
  \textbf{\bibinfo{volume}{19}}, \bibinfo{pages}{e071136}
  (\bibinfo{year}{2023}).

\bibitem{andreolettiMappingEthicalLandscape2024}
\bibinfo{author}{Andreoletti, M.}, \bibinfo{author}{Haller, L.},
  \bibinfo{author}{Vayena, E.} \& \bibinfo{author}{Blasimme, A.}
\newblock \bibinfo{title}{Mapping the ethical landscape of digital biomarkers:
  {{A}} scoping review}.
\newblock \emph{\bibinfo{journal}{PLOS Digital Health}}
  \textbf{\bibinfo{volume}{3}}, \bibinfo{pages}{e0000519}
  (\bibinfo{year}{2024}).

\bibitem{maciasalonsoDefinitionsDigitalBiomarkers2024}
\bibinfo{author}{Macias~Alonso, A.~K.}, \bibinfo{author}{Hirt, J.},
  \bibinfo{author}{Woelfle, T.}, \bibinfo{author}{Janiaud, P.} \&
  \bibinfo{author}{Hemkens, L.~G.}
\newblock \bibinfo{title}{Definitions of digital biomarkers: A systematic
  mapping of the biomedical literature}.
\newblock \emph{\bibinfo{journal}{BMJ Health \& Care Informatics}}
  \textbf{\bibinfo{volume}{31}}, \bibinfo{pages}{e100914}
  (\bibinfo{year}{2024}).

\bibitem{montagBlurryBoundariesWhen2021}
\bibinfo{author}{Montag, C.}, \bibinfo{author}{Elhai, J.~D.} \&
  \bibinfo{author}{Dagum, P.}
\newblock \bibinfo{title}{On {{Blurry Boundaries When Defining Digital
  Biomarkers}}: {{How Much Biology Needs}} to {{Be}} in a {{Digital
  Biomarker}}?}
\newblock \emph{\bibinfo{journal}{Frontiers in Psychiatry}}
  \textbf{\bibinfo{volume}{12}}, \bibinfo{pages}{740292}
  (\bibinfo{year}{2021}).

\bibitem{mulinariAligningDigitalBiomarker2024}
\bibinfo{author}{Mulinari, S.}
\newblock \bibinfo{title}{Aligning digital biomarker definitions in psychiatry
  with the {{National Institute}} of {{Mental Health Research Domain Criteria}}
  framework}.
\newblock \emph{\bibinfo{journal}{NPP---Digital Psychiatry and Neuroscience}}
  \textbf{\bibinfo{volume}{2}}, \bibinfo{pages}{15} (\bibinfo{year}{2024}).

\bibitem{vasudevanDigitalBiomarkersConvergence2022}
\bibinfo{author}{Vasudevan, S.}, \bibinfo{author}{Saha, A.},
  \bibinfo{author}{Tarver, M.~E.} \& \bibinfo{author}{Patel, B.}
\newblock \bibinfo{title}{Digital biomarkers: {{Convergence}} of digital health
  technologies and biomarkers}.
\newblock \emph{\bibinfo{journal}{npj Digital Medicine}}
  \textbf{\bibinfo{volume}{5}}, \bibinfo{pages}{36} (\bibinfo{year}{2022}).

\bibitem{biomarkersdefinitionworkinggroupBiomarkersSurrogateEndpoints2001}
\bibinfo{author}{Group, B. D.~W.}
\newblock \bibinfo{title}{Biomarkers and surrogate endpoints: {{Preferred}}
  definitions and conceptual framework}.
\newblock \emph{\bibinfo{journal}{Clinical Pharmacology \& Therapeutics}}
  \textbf{\bibinfo{volume}{69}}, \bibinfo{pages}{89--95}
  (\bibinfo{year}{2001}).

\bibitem{inter-organizationprogrammeforthesoundmanagementofchemicalsBiomarkersRiskAssessment2001}
\bibinfo{editor}{{Inter-Organization programme for the sound management of
  chemicals}} (ed.) \emph{\bibinfo{title}{Biomarkers in Risk Assessment:
  Validity and Validation}} No. \bibinfo{number}{222} in
  \bibinfo{series}{Environmental Health Criteria} (\bibinfo{publisher}{World
  health organization}, \bibinfo{address}{Geneva}, \bibinfo{year}{2001}).

\bibitem{strimbuWhatAreBiomarkers2010}
\bibinfo{author}{Strimbu, K.} \& \bibinfo{author}{Tavel, J.~A.}
\newblock \bibinfo{title}{What are biomarkers?:}.
\newblock \emph{\bibinfo{journal}{Current Opinion in HIV and AIDS}}
  \textbf{\bibinfo{volume}{5}}, \bibinfo{pages}{463--466}
  (\bibinfo{year}{2010}).

\bibitem{hartlTranslationalPrecisionMedicine2021}
\bibinfo{author}{Hartl, D.} \emph{et~al.}
\newblock \bibinfo{title}{Translational precision medicine: An industry
  perspective}.
\newblock \emph{\bibinfo{journal}{Journal of Translational Medicine}}
  \textbf{\bibinfo{volume}{19}}, \bibinfo{pages}{245} (\bibinfo{year}{2021}).

\bibitem{dorseyFirstFrontierDigital2017}
\bibinfo{author}{Dorsey, E.~R.}, \bibinfo{author}{Papapetropoulos, S.},
  \bibinfo{author}{Xiong, M.} \& \bibinfo{author}{Kieburtz, K.}
\newblock \bibinfo{title}{The {{First Frontier}}: {{Digital Biomarkers}} for
  {{Neurodegenerative Disorders}}}.
\newblock \emph{\bibinfo{journal}{Digital Biomarkers}}
  \textbf{\bibinfo{volume}{1}}, \bibinfo{pages}{6--13} (\bibinfo{year}{2017}).

\bibitem{wachterRightReasonableInferences2019}
\bibinfo{author}{Wachter, S.} \& \bibinfo{author}{Mittelstadt, B.}
\newblock \bibinfo{title}{A {{Right}} to {{Reasonable Inferences}}}.
\newblock \emph{\bibinfo{journal}{Columbia Business Law Review}}
  \bibinfo{pages}{494--620 Pages} (\bibinfo{year}{2019}).

\bibitem{vayenaStrictlyBiomedicalSketching2016}
\bibinfo{author}{Vayena, E.} \& \bibinfo{author}{Gasser, U.}
\newblock \bibinfo{title}{ in \textit{``{{Strictly Biomedical}}? {{Sketching}}
  the {{Ethics}} of the {{Big Data Ecosystem}} in {{Biomedicine}}''}} (eds
  \bibinfo{editor}{Mittelstadt, B.~D.} \& \bibinfo{editor}{Floridi, L.})
  \emph{\bibinfo{booktitle}{The {{Ethics}} of {{Biomedical Big Data}}}},
  Vol.~\bibinfo{volume}{29} \bibinfo{pages}{17--39}
  (\bibinfo{publisher}{Springer International Publishing},
  \bibinfo{address}{Cham}, \bibinfo{year}{2016}).

\bibitem{ruckensteinDataficationHealth2017}
\bibinfo{author}{Ruckenstein, M.} \& \bibinfo{author}{Sch{\"u}ll, N.~D.}
\newblock \bibinfo{title}{The {{Datafication}} of {{Health}}}.
\newblock \emph{\bibinfo{journal}{Annual Review of Anthropology}}
  \textbf{\bibinfo{volume}{46}}, \bibinfo{pages}{261--278}
  (\bibinfo{year}{2017}).

\bibitem{vandijckDataficationDataismDataveillance2014}
\bibinfo{author}{Van~Dijck, J.}
\newblock \bibinfo{title}{Datafication, dataism and dataveillance: {{Big Data}}
  between scientific paradigm and ideology}.
\newblock \emph{\bibinfo{journal}{Surveillance \& Society}}
  \textbf{\bibinfo{volume}{12}}, \bibinfo{pages}{197--208}
  (\bibinfo{year}{2014}).

\bibitem{mejiasDatafication2019}
\bibinfo{author}{Mejias, U.~A.} \& \bibinfo{author}{Couldry, N.}
\newblock \bibinfo{title}{Datafication}.
\newblock \emph{\bibinfo{journal}{Internet Policy Review}}
  \textbf{\bibinfo{volume}{8}} (\bibinfo{year}{2019}).

\bibitem{hoeyerDataficationAccountabilityPublic2019}
\bibinfo{author}{Hoeyer, K.}, \bibinfo{author}{Bauer, S.} \&
  \bibinfo{author}{Pickersgill, M.}
\newblock \bibinfo{title}{Datafication and accountability in public health:
  {{Introduction}} to a special issue}.
\newblock \emph{\bibinfo{journal}{Social Studies of Science}}
  \textbf{\bibinfo{volume}{49}}, \bibinfo{pages}{459--475}
  (\bibinfo{year}{2019}).

\bibitem{luptonCriticalDigitalHealth2016}
\bibinfo{author}{Lupton, D.}
\newblock \bibinfo{title}{Towards critical digital health studies:
  {{Reflections}} on two decades of research in {\emph{health}} and the way
  forward}.
\newblock \emph{\bibinfo{journal}{Health: An Interdisciplinary Journal for the
  Social Study of Health, Illness and Medicine}} \textbf{\bibinfo{volume}{20}},
  \bibinfo{pages}{49--61} (\bibinfo{year}{2016}).

\bibitem{mittelstadtEthicsBigData2016}
\bibinfo{author}{Mittelstadt, B.~D.} \& \bibinfo{author}{Floridi, L.}
\newblock \bibinfo{title}{ in \textit{The {{Ethics}} of {{Big Data}}:
  {{Current}} and {{Foreseeable Issues}} in {{Biomedical Contexts}}}} (eds
  \bibinfo{editor}{Mittelstadt, B.~D.} \& \bibinfo{editor}{Floridi, L.})
  \emph{\bibinfo{booktitle}{The {{Ethics}} of {{Biomedical Big Data}}}},
  Vol.~\bibinfo{volume}{29} \bibinfo{pages}{445--480}
  (\bibinfo{publisher}{Springer International Publishing},
  \bibinfo{address}{Cham}, \bibinfo{year}{2016}).

\bibitem{hofmannManagingMoralExpansion2022}
\bibinfo{author}{Hofmann, B.}
\newblock \bibinfo{title}{Managing the moral expansion of medicine}.
\newblock \emph{\bibinfo{journal}{BMC Medical Ethics}}
  \textbf{\bibinfo{volume}{23}}, \bibinfo{pages}{97} (\bibinfo{year}{2022}).

\bibitem{hofmannKairosDiagnostics2024}
\bibinfo{author}{Hofmann, B.} \& \bibinfo{author}{Wiesing, U.}
\newblock \bibinfo{title}{Kairos in diagnostics}.
\newblock \emph{\bibinfo{journal}{Theoretical Medicine and Bioethics}}
  \textbf{\bibinfo{volume}{45}}, \bibinfo{pages}{99--108}
  (\bibinfo{year}{2024}).

\bibitem{roseBiologicalCitizenship2005}
\bibinfo{author}{Rose, N.~S.} \& \bibinfo{author}{Novas, C.}
\newblock \bibinfo{title}{ in \textit{Biological {{Citizenship}}}} (eds
  \bibinfo{editor}{Ong, A.} \& \bibinfo{editor}{Collier, S.})
  \emph{\bibinfo{booktitle}{Global {{Assemblages}}: {{Technology}},
  {{Politics}}, and {{Ethics}} as {{Anthropological Problems}}}}
  \bibinfo{pages}{439--463} (\bibinfo{publisher}{Blackwell Publishing},
  \bibinfo{address}{Oxford}, \bibinfo{year}{2005}).

\bibitem{rosePoliticsLifeItself2009}
\bibinfo{author}{Rose, N.~S.}
\newblock \emph{\bibinfo{title}{The {{Politics}} of {{Life Itself}}:
  {{Biomedicine}}, {{Power}}, and {{Subjectivity}} in the {{Twenty-First
  Century}}}} In-{{Formation}} (\bibinfo{publisher}{Princeton University
  Press}, \bibinfo{address}{Princeton}, \bibinfo{year}{2009}).

\bibitem{brownMoralResponsibilisationHealth2019}
\bibinfo{author}{Brown, R. C.~H.}, \bibinfo{author}{Maslen, H.} \&
  \bibinfo{author}{Savulescu, J.}
\newblock \bibinfo{title}{Against {{Moral Responsibilisation}} of {{Health}}:
  {{Prudential Responsibility}} and {{Health Promotion}}}.
\newblock \emph{\bibinfo{journal}{Public Health Ethics}}
  \textbf{\bibinfo{volume}{12}}, \bibinfo{pages}{114--129}
  (\bibinfo{year}{2019}).

\bibitem{ranischFoundationModelsMedicine2025}
\bibinfo{author}{Ranisch, R.} \& \bibinfo{author}{Haltaufderheide, J.}
\newblock \bibinfo{title}{Foundation models in medicine are a social
  experiment: Time for an ethical framework}.
\newblock \emph{\bibinfo{journal}{npj Digital Medicine}}
  \textbf{\bibinfo{volume}{8}}, \bibinfo{pages}{525} (\bibinfo{year}{2025}).

\end{thebibliography}

\end{document}